\documentclass[lettersize,journal]{IEEEtran}
\usepackage{bbding}
\usepackage{multirow}

\usepackage{setspace}
\usepackage{cite,graphicx,amsmath,amssymb}
\usepackage{fancyhdr}
\usepackage{mdwmath}
\usepackage{mdwtab}
\usepackage{balance}
\usepackage{xcolor}
\usepackage{bm}
\usepackage{amsthm}
\usepackage{threeparttable}
\usepackage{algorithm}
\usepackage{algorithmic}
\usepackage{multirow}
\usepackage{flafter}
\usepackage{makecell}
\usepackage{diagbox}
\usepackage{caption}
\usepackage{subcaption}
\usepackage{ulem}
\usepackage{stfloats}
\usepackage{hyperref}

\providecommand{\url}[1]{#1}
\allowdisplaybreaks
\begin{document}

\vspace{-3em}
\title{Generative Artificial Intelligence (GAI) for Mobile Communications: A Diffusion Model Perspective}

\author{
  Xiaoxia Xu, 
  Xidong Mu, 
  Yuanwei Liu, \IEEEmembership{Fellow,~IEEE,} \\
  Hong Xing, 
  Yue Liu, 
  Arumugam Nallanathan, \IEEEmembership{Fellow,~IEEE}
  \vspace{-2em}

\thanks{Xiaoxia Xu is with the School of Electronic Engineering and Computer Science, Queen Mary University of
London, London E1 4NS, U.K. (email: x.xiaoxia@qmul.ac.uk).} 
\thanks{Xidong Mu is with the Centre for Wireless Innovation (CWI), Queen's University Belfast, Belfast, BT3 9DT, U.K. (e-mail: x.mu@qub.ac.uk).}
\thanks{
Yuanwei Liu and Arumugam Nallanathan are with the School of Electronic Engineering and Computer Science, Queen Mary University of
London, London E1 4NS, U.K., and also with the Department of Electronic Engineering, Kyung Hee University, Yongin-si, Gyeonggi-do 17104, Korea.
(email: yuanwei.liu@qmul.ac.uk, a.nallanathan@qmul.ac.uk). 
}
\thanks{Hong Xing is with the IoT Thrust, The Hong Kong University of Science and Technology (Guangzhou), Guangzhou, 511453, China; 
she is also affiliated with the Department of ECE, The Hong Kong University of Science and Technology,
HK SAR (e-mail: hongxing@ust.hk).}
\thanks{Yue Liu is with the Faculty of Applied Sciences, Macao Polytechnic University, Macau, SAR, China (e-mail: yue.liu@mpu.edu.mo).}
}

\maketitle

\begin{abstract}  
  This article targets at unlocking the potentials of a class of prominent generative artificial intelligence (GAI) method, namely diffusion model (DM), for mobile communications. 
  First, a DM-driven communication architecture is proposed, which introduces two key paradigms, i.e., conditional DM and DM-driven deep reinforcement learning (DRL), 
  for wireless data generation and communication management, respectively.
  Then, we discuss the key advantages of DM-driven communication paradigms. 
  To elaborate further, we explore DM-driven channel generation mechanisms for channel estimation, extrapolation, and feedback in multiple-input multiple-output (MIMO) systems. 
  We showcase the numerical performance of conditional DM using the accurate DeepMIMO channel datasets, 
  revealing its superiority in generating high-fidelity channels and mitigating unforeseen distribution shifts in sophisticated scenes.
  Furthermore, several DM-driven communication management designs are conceived, 
  which is promising to deal with imperfect channels and task-oriented communications. 
  To inspire future research developments, we highlight the potential applications and open research challenges of DM-driven communications. 
  Code is available at \url{https://github.com/xiaoxiaxusummer/GAI_COMM/}
\end{abstract}

\begin{IEEEkeywords}
Diffusion model, generative artificial intelligence (GAI), machine learning, mobile communications.
\end{IEEEkeywords}

\section{Introduction}
With recent strides in generative artificial intelligence (GAI), 
diffusion models (DM) has risen as a distinguished category of learning techniques \cite{Diffusion_Vision}, 
which present unparalleled content generation performance at human level and even beyond. 
Before the introduction of GAI, conventional discriminative AI typically estimates wireless data by 
minimizing average distances between predicted samples and ground-truth samples (such as cross entropy or mean square error (MSE)),
which lacks the modelling of the entire data distribution. 
As a typical GAI method, generative adversarial networks (GANs) jointly train generative and discriminative models by adversarial learning.  
However, the generator may focus on cheating the discriminator rather than capturing accurate data distributions, leading to \textit{mode collapse} and unstable training.
As another representative GAI approach, variational auto-encoder (VAE) learns low-dimension Gaussian data distribution over the latent space, 
which still suffers from distorted sampling due to the balance between regularization and reconstruction losses and the Gaussian data distribution assumptions.
Unlike previous data-driven learning methods, 
DM generates high-fidelity data from a stochastic step-by-step denoising process,  
which can accurately approximate complex data distributions in dynamic environment robust to unseen perturbations.
Hence, DM provides exceptional representation capability, diversity, and stability, which has shown state-of-the-art AI generated content (AIGC) performance 
in various challenging tasks such as text-to-image generation, image inpainting, and image super-resolution\cite{Score_SDE,Latent_diffusion}. 

While DM has become popular AI tools in natural language processing and computer vision domains, their applications in mobile communications are still at an early stage. 
Specifically, a denoising diffusion probabilistic model (DDPM) based transceiver design was proposed in \cite{DDPM_PHY_Impair}  for hardware-impaired communications. 
For near-field integrated sensing and communication (ISAC), the authors in \cite{Diffusion_ISAC} demonstrated the effectivity of DM for direction of arrival (DoA) estimation.
Moreover, several applications of DM in solving network optimization tasks were showcased in \cite{BeyondDRL_DM_NetworkOpt}. 
Nevertheless, there is still a limited understanding of how DM can introduce new perspectives and enable performance gains in mobile communications. 
Furthermore, effective means for fully harnessing DM techniques in physical layer designs and network resource management are yet to be explored, 
particularly for future mobile networks relying on massive antennas and experiencing highly dynamic environment. 
To unlock the potentials of DM techniques and pave the way forward, 
this article aims to present an overview of DM-driven mobile communications and delve in how they can surpass conventional AI paradigms. 
Focusing on wireless data generation and dynamic resource management, 
meaningful and reliable DM-driven AI mechanisms can be achieved for mobile communications with intricate stochastic perturbations and possible distribution shifts.  
The main contributions of this article can be summarized as follows.

\begin{itemize}
\item We propose a DM-driven communication architecture, which involves two main categories of paradigms, namely conditional DM and DM-driven RL. 
We identify the significant roles of these DM-driven paradigms in mobile communications. 
Then, their fundamental principles, key advantages, and representative operating mechanisms are reviewed and discussed. 
\item We discuss efficient DM-driven wireless channel generation mechanisms for MIMO channel estimation, extrapolation, and feedback. 
Moreover, we provide case studies over DeepMIMO channels constructed by accurate ray-tracing data \cite{DeepMIMO}. 
Numerical results demonstrate that conditional DM significantly improves the channel estimation accuracy even in mixed scenes or in the presence of data distribution shifts.
\item We conceive DM-driven communication management designs. 
Several DM-driven RL schemes are explored to deal with imperfect channel state information (CSI) for block-based and end-to-end communication structures. 
Furthermore, a DM-driven meta-RL scheme is introduced for task-oriented communications, which can identify and generalize across multiple communication task contexts. 
\item We outline the promising applications of DM-driven communications in unmanned aerial vehicles (UAVs)-assisted networks and ISAC systems.
Furthermore, we highlight open challenges of deploying DM in future mobile communication systems. 
\end{itemize}

\begin{figure*}[!t]
  \vspace{-2em}
  \centering
  \includegraphics[width=0.8\linewidth]{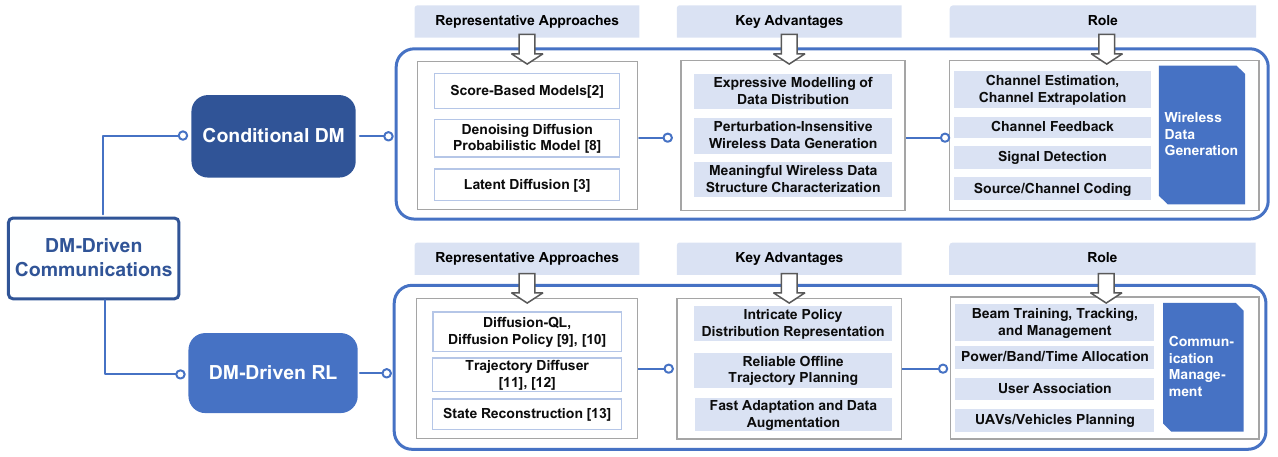}
  \caption{An overview of DM-driven communications.}
  \vspace{-1em}
\end{figure*}


\vspace{-0.5em}
\section{An Overview of DM-Driven Communications}
 
This section presents an overview of DM-driven communications. 
As shown in Fig. 1, we broadly categorize DM-driven communication methods into two key paradigms, referred to as conditional DM and DM-driven reinforcement learning (RL), respectively. 
These paradigms empower smart mobile communications from two different perspectives. 
Particularly, conditional DM enables high-fidelity wireless data generation, whilst DM-driven RL empowering efficient resource management.
By exploiting different training/execution mechanisms, they can play distinct roles in mobile communications. 
In this section, we will review the basic principles of conditional DM and DM-driven RL, as well as identify their key advantages compared to conventional learning paradigms.

\begin{figure*}[!bp]
  \centering
  \includegraphics[width=0.78\linewidth]{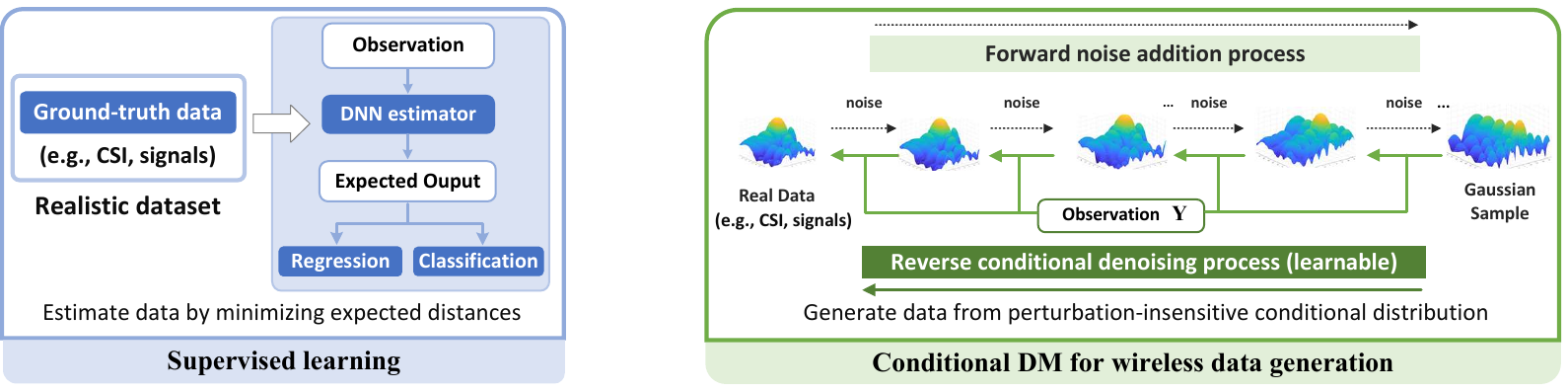}
  \caption{Learning mechanisms of conventional supervised learning and conditional DM for wireless data generation.}
  \label{fig_diffusion_dnn}
\end{figure*}

\subsection{Conditional DM: Principles and  Advantages}

Conditional DM can automatically generate wireless data (e.g., channel state information (CSI) and target wireless signals) 
conditioned on specific wireless system observations.
As shown in Fig. \ref{fig_diffusion_dnn}, conditional DM employs a disruptive learning and generation paradigm, 
which consists of a forward noise addition process and a reverse denoising process \cite{Score_SDE,DDPM}. 
Specifically, the forward noise addition process gradually adds random noises to the realistic data. 
These noises are sampled from a set of Gaussian distributions with different means and variances. 
Then, a conditional denoising model (parameterized by deep neural network (DNN)) is trained in the reverse denoising process 
to recover ground-truth samples from noise-corrupted data. 
During execution,  high-fidelity wireless data can be sampled from the learned distribution using the denoising model guided by observed environment conditions. 
The discrete-time conditional DM can be implemented by DDPM \cite{DDPM}, which models the entire process as a discrete Markov chain. 
To enhance the flexibility and representational capability, score-based stochastic differential equations (SDE) 
further generalizes the discrete-time denoising process to continuous time \cite{Score_SDE}. 
Different from step-by-step denoising sampling in DDPM, score-based models exploit Euler-Maruyama predictor to approximate reverse SDE solutions for denoising.
To reduce noise prediction errors, Markov Chain Monte Carlo (MCMC), e.g., Langevin dynamics, can be combined to correct the noise predictor, known as predictor-corrector sampling.

Generally, conditional DM enables an innovative paradigm of wireless data generation for mobile communications, such as CSI estimation, channel coding/decoding, 
and signal detection. 
The key advantages are summarized as follows. 
\begin{itemize}
  \item{\textbf{Expressive modelling of wireless data distribution:}}
Condition DM aims to approximate the full data distribution from a statistical perspective.
  Unlike supervised learning that minimizes average distances, or GANs that rely on adversarial training, it optimizes statistical objectives such as Kullback-Leibler (KL) divergence. 
  This enables a more accurate and complete representation of diverse patterns existing in wireless data, 
  thus providing an expressive modelling of sophisticated data distributions in mobile communications.
  \item{\textbf{Perturbation-insensitive wireless data generation:}}
  DM learns to recover wireless data from a severely distorted distribution. 
  Different from GAN and VAE that employ a deterministic discriminator or encoder, 
  conditional DM is resistant to mixed disturbances from multi-level additive noises.
  This is aligned with the stochastic and noise-prone natures of wireless communications, 
  which enables conditional DM to generate wireless data in a perturbation-insensitive and noise-resistant way. 
  \item{\textbf{Meaningful data structure characterization:}}
  Conventional data-driven methods may ignore specific data structures and finer details.
  By training the reverse denoising model to reconstruct realistic data from a stochastic process, 
  conditional DM is forced to capture key features and intrinsic structures of wireless data. 
  The denoising model can identify the core patterns and detailed features of wireless data from meaningless noises, 
  which efficiently characterizes the underlying dependence of realistic wireless data (e.g., wireless signal correlations and channel spatial characteristics). 
\end{itemize}

\begin{figure*}[!tp]
  \vspace{-3em}
  \centering  
  \includegraphics[width=0.85\linewidth]{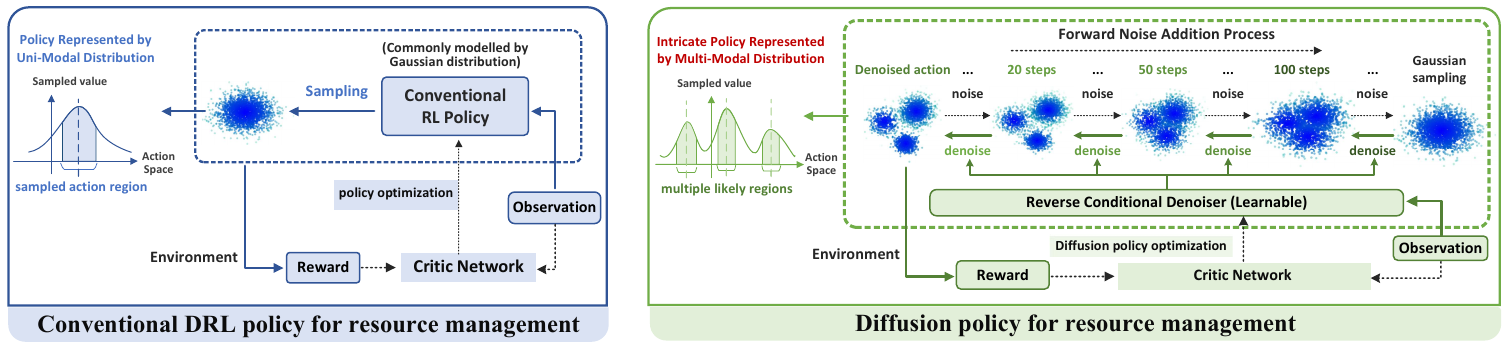}
  \caption{Learning mechanisms of conventional RL and the emerging DM-driven RL for communication management.}
  \label{fig_diffusion_architecture}
  \vspace{-1em}
\end{figure*}

\vspace{-0.5em}
\subsection{DM-Driven RL: Principles and Advantages}
While conditional DM targeting at accurately reconstructing the ground-truth data, 
the goal of DM-driven RL is to explore better solutions without knowing the realistically optimal one. 
Due to the ever-increasing system complexity, future mobile communications should deal with a variety of resource management problems, 
thus efficiently assigning multi-domain radio resources (such as RF components, bandwidth, and power) in a dynamic environment.
Due to the non-convexity of communication management, existing expert-based decision-making strategies are usually sub-optimal.
Following the basic principle of deep reinforcement learning, DM-driven RL can explore and generate decisions via the trial-and-error mechanism. 
Fig. \ref{fig_diffusion_architecture} demonstrates a representative DM-driven RL paradigm, namely diffusion policy, which models the RL policy as a reverse denoising diffusion model.  
In addition to the policy representation, DM can also model trajectory dynamics or reconstruct out-of-distribution (OOD) states for RL. 
Compared to conventional deep RL, the key advantages of DM-driven RL can be analyzed as follows.
\begin{itemize}
  \item \textbf{Intricate policy distribution representation and efficient exploration}: 
  Current RL algorithms typically parameterize policies by Gaussian distributions,  
  which only generates actions around the predicted mean value, leading to a uni-modal distribution of policy representation, as shown in Fig. \ref{fig_diffusion_architecture}. 
  In contrast, diffusion policy can generate actions within multiple likely regions  
  by denoising the uni-modal Gaussian distribution into a multi-modal policy distribution \cite{Diffusion_OnlineRL,Diffusion_OfflineRL}. 
  Therefore, DM can approximate intricate policy distributions that combines multiple potential decisions for various application scenarios. 
  This can also improve RL's policy exploration efficiency to rapidly discover useful solutions by interacting with environment.
\item \textbf{Trajectory-based planning for reliable offline RL:}  
  Since exploring new policies in real-world environment may lead to high-risk behaviours, 
  for many risk-sensitive mobile applications (e.g. Internet of Vehicles), RL policies can only be trained using historically collected datasets. 
  This leads to the concept of offline RL that learns decision-making policy without online interactions with environment. 
  DM can accurately model and predict the environment dynamics from historical trajectories, 
  thus enabling trajectory-level reliable action planning in offline RL \cite{Diffusion_OfflineRL_Trajectory_Planning}. 
  Moreover, offline RL methods are generally susceptible to states/actions outside the data distribution of previously collected datasets, 
  and are prone to performance crashes due to distribution shifts.
  As a remedy, DM can synthesize different tasks' experiences for conditional trajectory generation, 
  thus generalizing across unseen communication tasks as a meta-diffuser \cite{MetaDiffuser}. 
  Further, DM can also be leveraged to reconstruct OOD states for robust decision-making \cite{Diffusion_OfflineRL_State}.  
\end{itemize}

\section{DM-Driven Channel Generation}

In this section, we discuss promising conditional DM mechanisms for MIMO channel generation, thus efficiently tackling channel estimation, channel extrapolation, and channel feedback. 
Furthermore, case studies over DeepMIMO channels in various scenes are provided for numerical evaluation.

\subsection{Conditional DM Mechanisms}

First, we discuss effective DM mechanisms to realize conditional data generation for MIMO channel estimation, extrapolation, and feedback.
\subsubsection{Channel Estimation} 
Conventional channel estimation (e.g., compressed sensing (CS)) relies on channel sparsity and prior knowledges of antenna array configurations. 
As a data-driven approach, GAI-assisted channel estimation does not rely on specific assumptions. 
Existing GAN-based channel estimation methods have demonstrated exceptional performance in modelling channel distributions over specific datasets. 
As an advance, conditional DM can learn the intrinsic structure of MIMO channels in various real-world scenes,
thus reconstructing realistic channels according to received pilot signals. 
Specifically, conditional DM enjoys several benefits:
(i) Enhance the robustness to different communication scenes and array configurations. (ii) Overcome physical layer impairments, such as noise, channel fading, and interference. 
(iii) Deal with hardware imperfections, such as mutual coupling, quantization errors and antenna position perturbations.

\begin{figure}[!bp]
  \centering
  \includegraphics[width=1\linewidth]{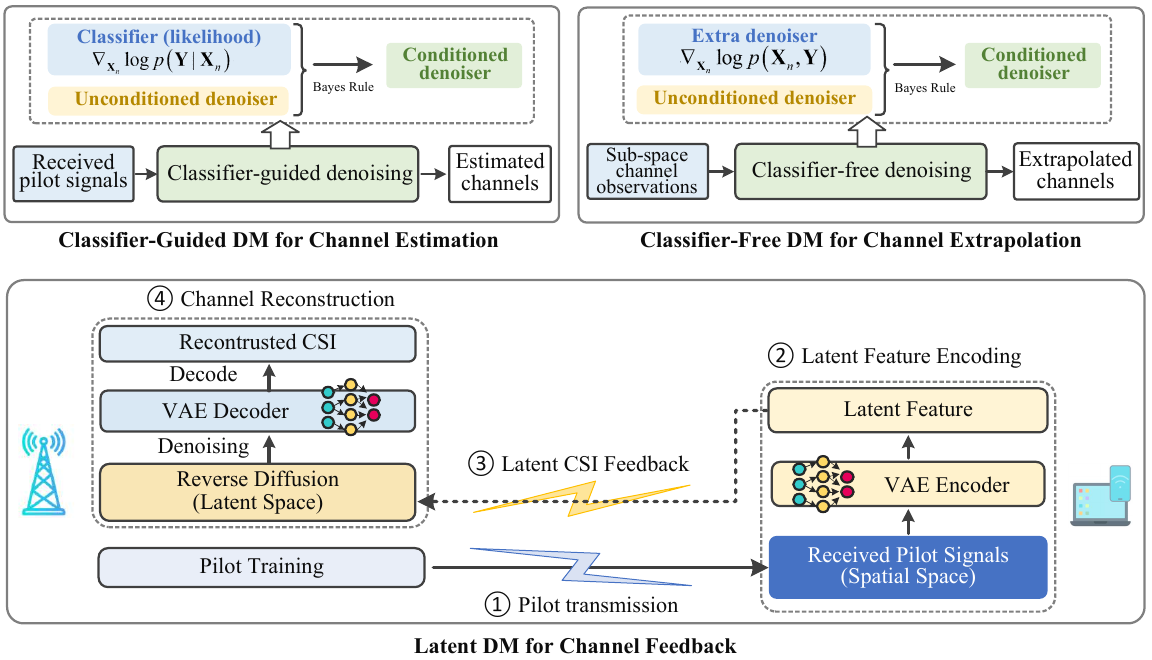}
  \caption{DM-driven wireless channel generation mechanisms.}
  \label{fig_channel_generation}
\end{figure}

To generate channels based on the received pilot signals, a critical problem is how to estimate the conditional distribution from an unconditioned denoising model to sample channels. 
As shown in Fig. \ref{fig_channel_generation}, the conditional channel generation can be realized by classifier-guided sampling.
Specifically, since the received pilot signals can be explicitly represented as a function of MIMO channels, 
the likelihood function of the received pilot signals given a channel sample can be estimated. 
Then, by combining the likelihood function with a pre-trained unconditioned channel denoiser, a conditioned denoiser can be estimated based on Bayes rule, 
which iteratively recovers CSI in the denoising process according to pilot signal observations.
As the likelihood function is also known as classifier in image generation, 
this posterior sampling mechanism is commonly referred to as classifier-guided conditional generation.

\subsubsection{Channel Extrapolation}
Channel extrapolation reconstructs the entire spatial channels  by sensing only a small portion of spatial channel parameters in massive MIMO systems. 
Specifically, the unknown sub-space channel parameters can be extrapolated from the sensed parameters over another subspace based on there underlying interdependence.
Channel extrapolation can be implemented over different antennas, frequencies, and terminals, respectively. 
The main difficulty is to model the mapping function from the sensed channel sub-space to the target sub-space. 
Since this mapping function behaves like a black box that lacks an explicit mathematical representation, it can be modelled by data-driven deep learning methods, as discussed in \cite{Channel_Extrapolation}. 
However, conventional data-driven learning (such as DNN and GAN) typically suffers from limited generalization capability, 
sensitivity to noises in data inputs, and susceptibility to OOD data in mobile environment.
In comparison, by learning from stochastic generation process with step-wise denoising, 
conditional DM can effectively capture a broader range of channel distribution patterns, deal with noisy/imperfect observations, and improve the adaptability to channel distribution shifts. 
Thus, DM empowers more accurate extrapolation of high-dimension channel parameters for large-scale MIMO. 
It is worth noting that channel extrapolation requires a distinct conditional generation mechanism from channel estimation. 
Since the mapping function between channel subspaces is unknown, there is no prior-known classifier model for classifier-guided conditional generation. 
Hence, an independent classifier model should be pre-trained by discriminative AI models, which may introduce additional estimation losses. 
To mitigate this, classifier-free conditional DM mechanism  \cite{Classifier_Free_DM} can be further exploited, which introduces an extra denoiser, as shown in Fig. \ref{fig_channel_generation}. 
The extra denoiser and the unconditioned denoiser approximate the likelihood function and the conditional distribution based on the conditional probability definition and Bayes rule, respectively, 
and they can use shared parameters to reduce the computational complexity. 
By jointly training these two denoisers, high-fidelity channels can be sampled with pure generative models to enhance the estimation performance.

\subsubsection{Channel Feedback} 
The channel reciprocity cannot be guaranteed in frequency-division duplex (FDD) massive MIMO systems. 
Therefore, the high-dimensional downlink channel parameters should be estimated by the terminal device (TD) and then fed back to the base station (BS). 
This results in a fundamental trade-off between CSI feedback overheads and quality. 
To address this challenge, latent DM \cite{Latent_diffusion} can be invoked to compress CSI feedback into latent features, 
thus improving the estimation accuracy at reduced communication and computation overheads.
As indicated by Fig. \ref{fig_channel_generation}, the latent DM involves a pre-trained VAE for CSI feedback compression. 
The VAE encoder can be deployed at the TD to extract latent channel features from the received downlink pilot signals. 
The low-dimensional latent channel features are then transmitted from the TD to  the BS. 
Thereafter, BS can denoise the latent channel features via the reverse diffusion process, followed by which the spatial CSI is reconstructed using the VAE decoder. 
As latent DM performs denoising process in the low-dimensional latent space, it can significantly reduce the computational complexity. 
Furthermore, the diffusion denoiser and VAE encoder/decoder can be jointly trained to combat noisy wireless channels during transmitting the CSI feedback. 
By doing so, latent DM can support cost-effective and reliable channel feedback for FDD systems.

\subsection{Case Studies}

\begin{figure}[!tp]
  \centering
  \begin{subfigure}{0.8\linewidth}
    \centering
    \includegraphics[width=1\linewidth]{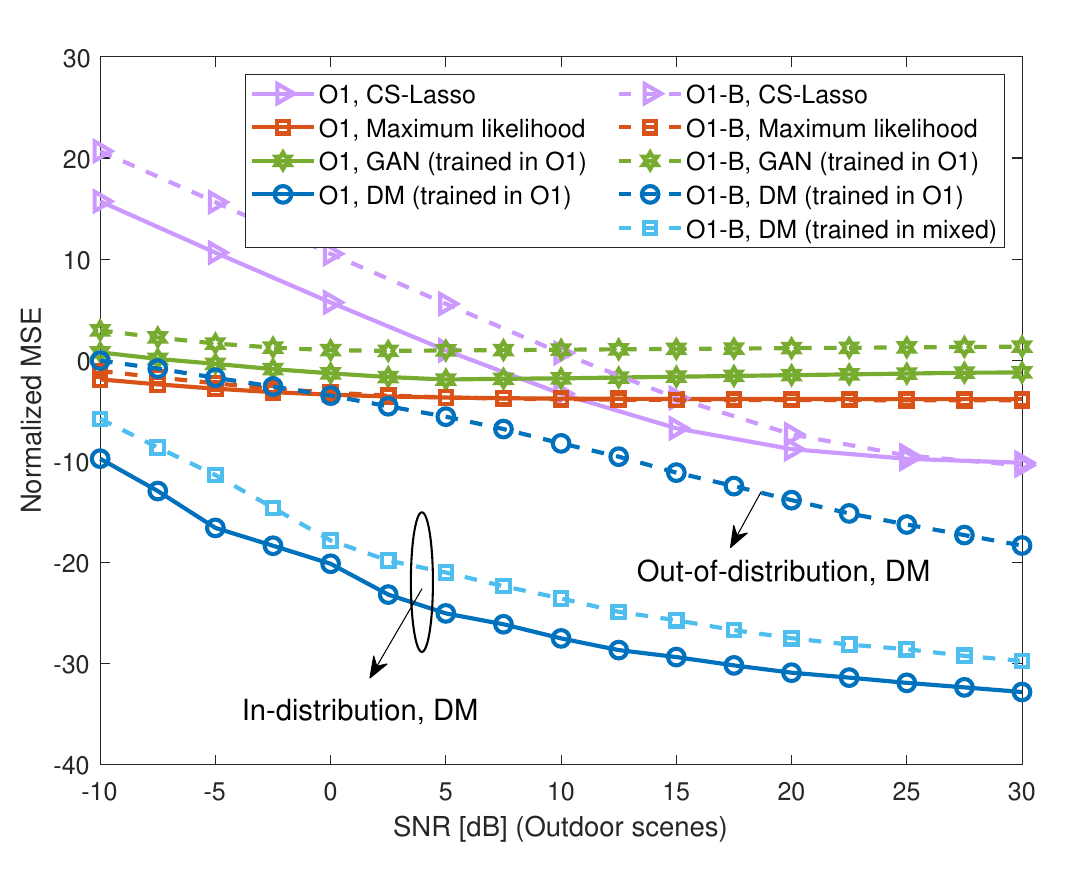}
    \caption{Performance comparisons in outdoor scene.}
  \end{subfigure}
  \\
  \begin{subfigure}{0.8\linewidth}
    \centering
    \includegraphics[width=1\linewidth]{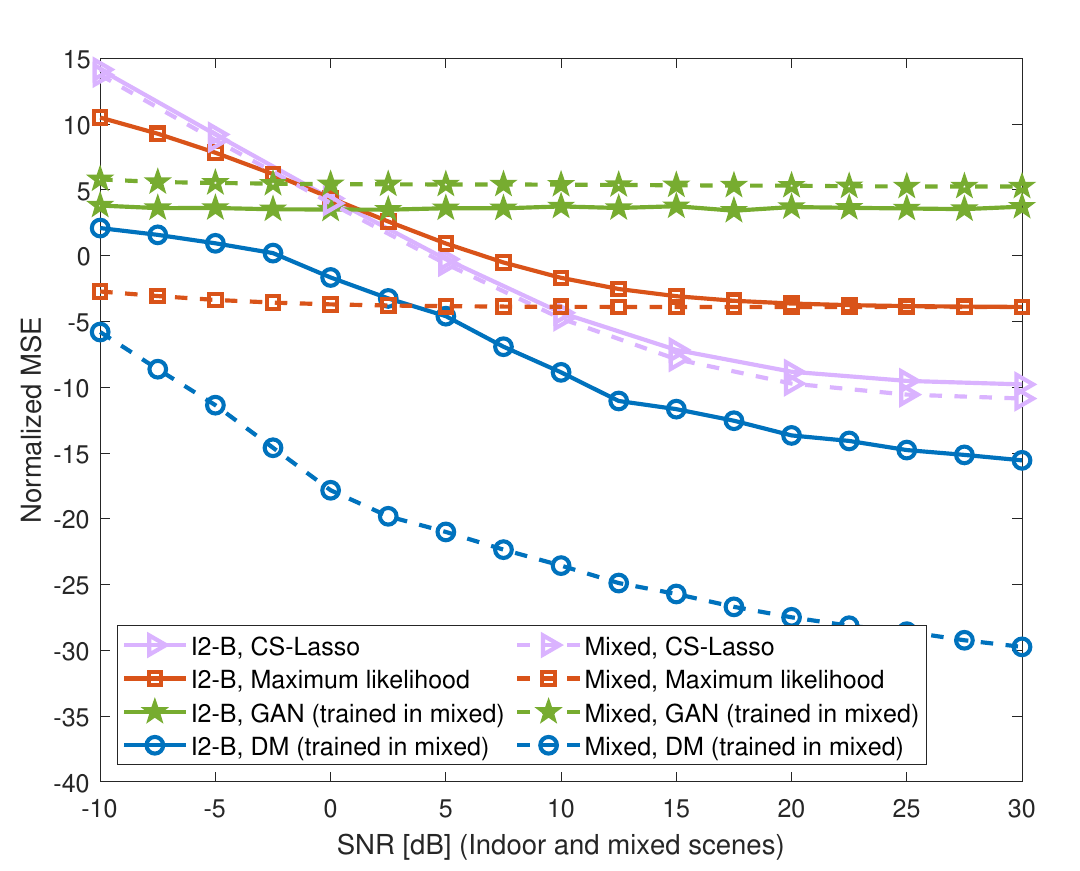}
    \caption{Performance comparisons in indoor and mixed scenes.}
  \end{subfigure}
  \caption{Channel estimation performance for $28$-GHz mmWave downlink channels based on DeepMIMO datasets in various scenes.}
  \label{fig_diffusion_CE}
\end{figure}

We evaluate the channel estimation performance of conditional DM through simulations over DeepMIMO datasets \cite{DeepMIMO}. 
Constructed by ray-tracing data, DeepMIMO datasets provide simulations of realistic MIMO channels in diverse scenarios and are aligned with 5G new radio (NR) standard. 
Specifically, we consider two outdoor scenes ``O1'' and ``O1-B'', one indoor blockage scene ``I2-B'', and a mixed scene of \{``O1'', ``O1-B'', ``I2-B''\}. 
Here, notation ``B'' indicates the existence of blockage. The carrier frequency for all these scenes is $28$ GHz. 
The BS and user are equipped with a $64$-antenna uniform planar array (UPA) and a $16$-antenna UPA, respectively. 
The number of transmitted pilot signals for channel estimation is $32$. 
We train conditional DM and conditional GAN in the ``O1'' and the mixed scenes, and then test them in different scenarios.  
Moreover, we implement classifier-guided conditional channel generation using predictor-corrector sampling. 
As shown in Fig. \ref{fig_diffusion_CE}, conditional DM outperforms conventional CS and conditional GAN in both outdoor and indoor scenes, 
and the performance gains increase significantly as the signal-to-noise ratio (SNR) increase. 
This is because conditional DM can fully capture MIMO channel structures and finer spatial features, and the conditional sampling becomes more accurate with less channel noises.
The conditional DM trained in the ``O1'' scene maintains a low MSE even when tested in the previously unseen ``O1-B'' scene, confirming its robustness to OOD blocking.
Notably, conditional DM dramatically improves the channel estimation accuracy in the mixed scenes, which demonstrates its potentials for representing complex channel distributions in realistic mobile communications.

\section{DM-Driven Communication Management}
This section explores efficient DM-driven communication management. 
We first conceive the DM-driven communication structures, which can deal with CSI and hardware imperfections for conventional block-based communication structure 
and the completely data-driven end-to-end communication structure. 
Thereafter, to mitigate simulation-to-reality gaps in more practical multi-task RL environment, we further conceive a DM-driven RL scheme for task-oriented communications.

\subsection{DM-Driven Communication Structure}
Mobile communication systems are witnessing an increasing hardware complexity with large-scale antennas. 
Therefore, acquiring accurate CSI may lead to overwhelming pilot overheads.
Moreover, as wireless systems evolve with time and user movement, predictive managements of transmitters and receivers are required to overcome outdated CSI 
and hardware impairments (such as power amplifier distortion and phase noises). 
DM offers an expressive AI method for proactive and dynamic communication management under CSI and hardware imperfections. 
In this part, we discuss tailored designs of DM-driven block-based and end-to-end communication structures.
\begin{itemize}
\item\textbf{DM-driven block-based communication structure}: 
The mobile communication process can be divided into multiple separate and manageable signal processing blocks, such as source coding, channel coding, modulation, beamforming, and channel estimation. 
DM can compensate for CSI and hardware imperfections by recovering inaccurate estimations from environment feedbacks. 
Moreover, to avoid error propagation of inaccurate and outdated CSI, 
RL policy can be customized to directly predict communication management solutions from the received pilot signals and the recovered features. 
To reduce computational complexity, cooperative denoising mechanism can be designed for block-based communication structure, whereby multiple signal processing blocks can share a part of denoising steps or intermediate denoising features. 
Furthermore, block-specific tiny DM models can be learnt from a large-scale DM by knowledge distillation, 
thus tailoring DM to different signal processing tasks and reducing model sizes based on real-time channel conditions. 
\item\textbf{DM-driven end-to-end communication structure}:
To mitigate inefficiencies of the conventional block-based communication structure,  DM can be exploited as a backbone module for end-to-end performance optimization. 
To characterize the physics of antenna hardware (e.g., antenna radiation directivity, polarization effects, 
and mutual coupling between antenna elements), and radio propagation (e.g., near-field/far-field communications), 
a probabilistic model can be incorporated into DM to capture latent representations of electromagnetic environment properties.
The physics-aware DM-driven end-to-end structure can be jointly trained to optimize multiple objectives, such as minimizing signal distortion and processing latency.
Based on this DM-driven end-to-end communication structure, physics-aware denoising can be performed for joint MIMO precoding and source-channel coding. 
To reduce the system complexity,  early-stopping denoising mechanism can be further developed according to real-time channel conditions.
\end{itemize}

\subsection{DM-Driven Task-Oriented Communications}
Practical mobile communications usually undergo a dynamic multi-task mixed environment, which consists of   
different task objectives (e.g., network and individual performance) and communication modelling (e.g., channel models and traffic distribution patterns). 
Therefore, it is essential to develop a task-oriented RL model, 
which handles a set of MDPs characterized by distinct reward functions and transition dynamics, 
as well as automatically generalize across different tasks and transfer from simulated environment to the reality. 
Existing online RL methods directly explore high-quality decision-making policies by interacting with environment, without understanding the underlying environment dynamics. 
Nevertheless, the online exploration is commonly error-prone and high-risk, which is not favourable to practical mobile network operations. 
Moreover, transferring online RL policies across multiple tasks usually requires an additional re-training process, which increases the costs for RL exploration.  
To overcome these shortcomings, a promising approach is to train RL policy offline, and then achieve simulation-to-reality transfer by online domain adaptation. 
Specifically, offline RL learns a task-oriented trajectory generation model to capture multi-task environment dynamics from historical datasets.  
Then, online transfer learning can be performed to identify distribution shifts and suppress the simulation-to-reality gap.
To reap the benefits, we introduce DM-driven  meta-RL for task-oriented communications as follows.
\begin{itemize}
  \item \textbf{Task context encoder pre-training}: 
  A context encoder is first pre-trained in mobile communication environment,  
  which identifies different task contexts by observing short fragments of multi-task trajectories from a mixed dataset. 
  By encoding trajectory fragments into a compact representation, it captures underlying dynamics of different communication task contexts. 
  \item \textbf{DM-driven offline meta-RL training}: Conditional DM can be invoked as a trajectory generator to predict action planning of mixed tasks conditioned on the predicted context. 
  The trajectory can be predicted by a reverse conditional diffusion process that mimics expert trajectories while improving accumulative returns. 
  During online execution, task contexts can be first identified from a small-size warm-start trajectory segment. 
  Thereafter, future trajectories are generated by DM to predict the optimal actions for decision-making over the planning horizon, thus acquiring long-horizon profits.
  \item \textbf{Online transfer learning}:
  Based on transfer learning, e.g., domain adaptation, 
  the simulation-to-reality model discrepancy can be mitigated by adjusting the trained models online to minimize impacts of distribution shifts.
\end{itemize}

\section{Applications and Challenges}

\subsection{Promising Applications}
The introduced DM-driven paradigms can empower wireless data generation and  resource management in existing mobile communication applications, 
including UAVs-assisted networks and ISAC, which can be outlined as follows.
\begin{itemize}
\item{\textbf{DM-driven UAVs-assisted networks}:}
UAVs can traverse complex terrains for fast-response and flexible network deployment in remote areas and disaster-stricken/emergency regions, 
thus compensating for the limited coverage of terrestrial networks. 
To realize ubiquitous and reliable connectivity,  UAVs-assisted networks require well-designed autonomous path planning, energy management, and multiagent collaboration strategies, 
which is challenging in high-mobility air-ground integrated communications. 
DM can predict the rapidly changing 3D spatial channels and potential blockages in different mobility patterns of UAVs.
Furthermore, as it is costly to collect training data for UAVs-assisted networks, diffusion policies can be explored in offline RL settings to achieve resilient and energy-efficient trajectory planning, spectrum sharing, and reliable mobile connectivity. 
\item{\textbf{DM-driven ISAC}:} 
Future mobile networks are expected to realize dual functionalities of high-precision radar sensing and high-quality mobile communications with unified hardware and software modules. 
Conditional DM can accurately recover sensing parameters and synthesize multi-view observations for radar sensing, e.g., for multi-target tracking and collaborative sensing. 
Furthermore, to achieve sensing-communication coexistence, DM-driven multi-task RL algorithm can be developed to generate dual-functional waveforms and 
configure dynamic resource management for task-oriented ISAC systems.
\end{itemize}

\subsection{Research Challenges and Opportunities}
To  deploy and exploit DM-driven communication paradigms in mobile communication networks, several open challenges remain to be addressed, which can be exemplified as follows. 
\begin{itemize}
  \item\textbf{DM sampling acceleration:}
    The reverse diffusion process necessitates tens to hundreds of denoising steps to ensure satisfactory performance. 
    Hence, it is critical to investigate efficient DM sampling acceleration methods, especially for mobile devices. 
    Specifically, tailored sampling techniques that incorporate higher-order information and neural ordinary differential equations (ODE) solvers can be investigated for wireless data generation. 
    Moreover, efficient distillation approach can be designed to adaptively reduce denoising steps and slim noise prediction networks in mobile applications. 
    \item\textbf{Mobile edge enabled DM:}
    The state-of-the-art DM relies on large-scale parameters and intensive computations (e.g., self-attention). 
    To reduce training overheads and minimize generation latency, 
    the radio/computing resources from both mobile edge servers and devices should be jointly exploited. 
    For the training of large-scale DM, communication-efficient and personalization-aware distributed learning mechanisms can be explored. 
    For collaborative and low-latency generation, denoising model split and offloading methods need to be investigated to efficiently utilize resources of hardware-limited devices.  
    \item\textbf{Interplays with model-driven learning:} 
    DM enables expressive data representations but requires high computing overheads. 
        On the other hand, model-driven learning approaches (e.g., deep unrolling) can leverage domain knowledges and mathematical optimization theory to guide black-box neural network designs, 
        thus improving the convergence speed and satisfying communication system constraints.
        This constitutes a research interest in combining DM with model-driven learning approaches, thus achieving cost-efficient data generation and constrained decision-making.
\end{itemize}

\section{Conclusion}
This article unveiled the potentials of emerging GAI techniques in mobile communications from a DM perspective.
To provide a general guideline, a DM-driven communication architecture has been put forward, 
which introduced two key paradigms, namely conditional DM and DM-driven RL, for expressive wireless data generation and efficient communicatio management, respectively. 
Efficient DM-driven wireless channel generation schemes have been explored for channel estimation, channel extrapolation, and channel feedback of massive MIMO systems. 
Then, numerical studies have been provided, which verified the superiority of conditional DM in representing mixed-scene channel distribution and overcoming potential distribution shifts. 
Furthermore, DM-driven communication management designs have been conceived to deal with imperfect CSI and task-oriented communications. 
For the future outlook, key applications and open research problems of DM-driven communications have been outlined.


\vspace{-1cm}

\begin{IEEEbiographynophoto} {Xiaoxia Xu} (Member, IEEE) is currently a Postdoctoral Researcher at Queen Mary University of London, U.K.  
  She received her B.Eng. and Ph.D. degree from Wuhan University in 2017 and 2023, respectively.
  She was also a visiting student with the Queen Mary University of London from 2021 to 2022. 
  Her research interests include NOMA, mmWave/THz MIMO, AI for wireless networks, and mobile edge generation.
\end{IEEEbiographynophoto}

\vspace{-1cm}

\begin{IEEEbiographynophoto} {Xidong Mu} (Member, IEEE) is currently a Lecturer at the Centre for Wireless Innovation (CWI), Queen's University Belfast, Belfast.  
  He received the IEEE ComSoc Outstanding Young Researcher Award for EMEA region in 2023. 
  He is the recipient of the 2024 IEEE Communications Society Heinrich Hertz Award, 
  the Best Paper Award in ISWCS 2022, the 2022 IEEE SPCC-TC Best Paper Award, and the Best Student Paper Award in IEEE VTC2022-Fall.
\end{IEEEbiographynophoto}

\vspace{-1cm}

\begin{IEEEbiographynophoto} {Yuanwei Liu} (Fellow, IEEE) is a (tenured) Full Professor at The University of Hong Kong and a visiting professor at Queen Mary University of London. 
  His research interests include NOMA, RIS/STARS, Integrated Sensing and Communications, Near-Field Communications, and 
  machine learning. He serves as a Co-Editor-in-Chief of IEEE ComSoc TC Newsletter, an Area Editor of IEEE TCOM and CL, and an Editor of the IEEE COMST/TWC/TVT/TNSE.  
\end{IEEEbiographynophoto}

\vspace{-1cm}

\begin{IEEEbiographynophoto} {Hong Xing} (Member, IEEE) received the B.Eng. degree in Electronic Sciences and Technologies from Zhejiang University, China, 
  and the Ph.D. degree in Wireless Communications from King's College London, U.K.. Since Jan. 2022, she has been an Assistant Professor with the IoT Thrust, 
  The Hong Kong University of Science and Technology (Guangzhou), and an Affiliate Assistant Professor with the Dept. of ECE, The Hong Kong University of Science and Technology. Her research interests include federated learning, mobile-edge computing, simultaneous localization and communication, and wireless information and power transfer. She received the Best 50 of IEEE Global Communications Conference (GLOBECOM’14) in 2014. 
She was an Associate Editor of IEEE ACCESS between Mar. 2019 and Mar. 2023, and is currently serving as an Editor of IEEE Wireless Communications Letters. 
\end{IEEEbiographynophoto}

\vspace{-1cm}

\begin{IEEEbiographynophoto} {Yue Liu} (Member, IEEE) received a B.Sc. degree from Beijing University of Posts and Telecommunications, Beijing, China, in 2010 and a Ph.D. degree in Electronic Engineering from Queen Mary University of London, London, U.K., in 2014. In the same year, she joined MPI-QMUL Information Systems Research Centre, Macao SAR, China as a researcher and then she has been working as a lecturer in the Faculty of Applied Sciences, Macao Polytechnic University, Macao SAR, China, since 2016. 
Her research interests include AI-driven Wireless Networks, 6G Networks, Machine Learning, Radio Resource Management (RRM) and RF Signal Sensing.
\end{IEEEbiographynophoto}

\vspace{-1cm}

\begin{IEEEbiographynophoto} {Arumugam Nallanathan} (Fellow, IEEE)  (a.nallanathan@qmul.ac.uk) 
is a professor and the head of the Communication Systems Research (CSR) group in Queen Mary University of London. 
His research interests include beyond 5G wireless networks, the Internet of Things, and AI for Wireless Communications.
\end{IEEEbiographynophoto}

\end{document}